\documentclass[tradiabstract]{aa}
\usepackage{natbib}
\usepackage{txfonts}
\input{psfig}

\newcommand{\HI}{{\ion{H}{1}}}

\newcommand{\kms}{$\,$km$\,$s$^{-1}$}
\newcommand{\ergs}{$\,$erg$\,$s$^{-1}$}

\newcommand{\msun}{{${\rm M}_\odot$}}
\newcommand{\msunyr}{{${\rm M}_\odot$ yr$^{-1}$}}

\newcommand{\co}{{CO(2-1)}}
\def\HI{H{\,\small I}}

\def\emph#1{{\sl #1}}
\newcommand{\ltsima} {$\; \buildrel < \over \sim \;$}
\newcommand{\gtsima} {$\; \buildrel > \over \sim \;$}
\newcommand{\lta} {\lower.5ex\hbox{\ltsima}}
\newcommand{\gta} {\lower.5ex\hbox{\gtsima}}

\input{psfig}
\begin{document}
\title{Tracing the extreme interplay between radio jets and the ISM in IC~5063}
\titlerunning{Feedback in action in IC~5063}
\authorrunning{Morganti et al.}
\author{Raffaella Morganti\inst{1,2}, Wilfred Frieswijk\inst{1}, Raymond J. B. Oonk\inst{1}, Tom Oosterloo\inst{1,2},  Clive Tadhunter \inst{3}}

\institute{Netherlands Institute for Radio Astronomy, Postbus 2,
7990 AA, Dwingeloo, The Netherlands
\and
Kapteyn Astronomical Institute, University of Groningen, Postbus 800,
9700 AV Groningen, The Netherlands
\and
Department of Physics and Astronomy, University of Sheffield, Sheffield, S7 3RH, UK
}
\offprints{morganti@astron.nl}

\date{Received ...; accepted ...}

\date{\today}

\abstract{We report the discovery with  the Atacama Pathfinder EXperiment (APEX) of an outflow of molecular gas  in the radio-loud Seyfert galaxy IC~5063 (z = 0.0110).  In addition to the  emission of the large-scale CO disk, a prominent blueshifted wing is observed in the CO(2-1) spectrum. IC~5063 represents one of the best cases of a fast jet-driven \HI\  (and ionized gas) outflow, which is located at the site of a radio-bright feature about 0.5 kpc from the nucleus.  It is possible that the blueshifted part of the molecular gas is associated with this outflow and  is accelerated by the interaction with the radio jet. The outflow of molecular gas is characterized by an H$_2$ mass of the outflowing component  of between  $2.25\pm 0.70 \times 10^7$ \msun\  and $1.29\pm 0.40  \times 10^8$ \msun\   and  a mass outflow rate  between $22$ and 129  \msunyr\  depending on the assumption for $\alpha_X$   and assuming a luminosity ratio L$^{\prime}$CO(2-1)/L$^{\prime}$CO(1-0) = 1. This  confirms that this may indeed be the dominant component in outflows driven by the nuclear activity that are  also found in other objects. However, this high mass outflow rate cannot easily be supported for a long time, suggesting that the gas outflow in IC~5063 happens in bursts and is in a particularly strong phase at present.  
Owing to its proximity, IC~5063 serves as an excellent laboratory for understanding the impact of radio jets on the gas-rich inter-stellar medium.  
}
\maketitle  

\section{Introduction}
\label{sec:introduction}


The mechanism of energy {\sl feedback} is considered  to be the key to successfully model galaxy evolution (Silk \& Rees 1998, di Matteo et al.\ 2005). In early-type galaxies, feedback is mainly provided by the energy released by the active nucleus, and gas outflows are one of the main signatures of this process in action  (see Fabian 2012). 

Outflows driven by nuclear activity can be initially launched from the accretion disk or dusty torus that surrounds the black hole (BH), in the form of radiatively driven winds.  
The other main mechanism often considered is acceleration of the gas via radio jets/lobes.  Understanding the relative importance of these mechanisms is the goal of many recent studies.  
Outflows have so far mainly been traced using ionized gas  (see e.g.\ Crenshaw, Kraemer \& George 2003,  Nesvadba et al.\ 2007, Holt et al.\ 2009,  Reeves et al. 2009, Tombesi et al. 2012, Harrison et al.\ 2012). Outflow effects have often been found to be confined to the innermost nuclear regions with  relatively modest associated mass outflow rates  (see e.g.\ Holt et al.\ 2011). Although this could have been a problem with the initial models of radiatively driven outflows (di Matteo et al. 2005), the possibility of a more complex, two-phase action on the inter-stellar medium (ISM) clouds makes this mechanism very relevant in the feedback process (Hopkins \& Elvis 2010).  

Interestingly, more  massive outflows  have recently been  discovered from the study of atomic neutral \HI\ and molecular gas  (e.g.\ Morganti et al.\ 2005a, 2005b, 2010;  Feruglio et al.\ 2010, Alatalo et al.\ 2011, Dasyra \& Combes 2011, 2012).  In a number of objects, radio plasma jets have been suggested to play the dominant role in driving the outflows.  Indeed, in some cases (Morganti et al. 2005b, Oosterloo et al. 2000), the location of the outflow appears to be clearly co-spatial with prominent radio features (e.g. hot spots).  This has also been independently confirmed by observations of the ionized gas component.  Radio jets can indeed provide a particularly suitable and fast way of transporting the energy because they couple efficiently to the ISM and produce fast  outflows from the central regions as required from feedback models (Wagner \& Bicknell  2011, Wagner, Bicknell, Umemura 2012). 
It is surprising, however, that despite the energetic interaction, these off-nuclear outflows can still have a component of  relatively cold gas ($< 1000$ K) as detected in \HI. 
Whether this is a common characteristic is an open question and is the motivation for the observations presented in this letter. If radio jets are as important - in certain conditions - as claimed by Wagner \& Bicknell  (2011) and Wagner et al. (2012),  the characteristics of the cold gas can provide constraints on the interaction physics.

\begin{figure}
\centerline{\psfig{file=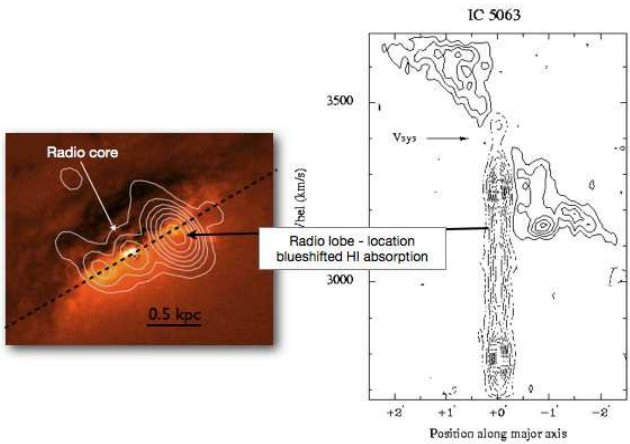,width=9cm,angle=0}}
\caption{Some relevant characteristics of IC~5063 (taken  from Morganti et al.\ 1998).  Left: the morphology of the radio continuum emission  (white contours)  is superimposed on an HST image. The contours show the bright radio lobe (the W component), the core, and the fainter E lobe. The APEX CO(2-1) observation is centered on the radio core but with a beam of 30$^{\prime\prime}$, it  covers the entire radio source. The right figure illustrates the \HI\ position-velocity plot taken along the major axis of the \HI\ disk (black dashed line on the radio image). The large-scale \HI\ disk in emission is shown with solid contours while the broad blueshifted component detected in absorption (dashed contours) is clearly evident, covering velocities well outside the range of the normal gas disk (Morganti et al. 1997). The spatial axis in this plot covers the galaxy out to 2.5 arcmin radius. VLBI follow-up \HI\ observations have shown that the blueshifted absorption is located at the position of the western brighter radio lobe, about 0.5 kpc from the nucleus, suggesting that the jet-ISM interaction occurs  at this location (Oosterloo et al. 2000).}
\label{fig:panel}
\end{figure}

Here we present the results of CO(2-1) observations of the nearby, southern radio-loud Seyfert galaxy IC~5063 ($z = 0.0110$)\footnote{Throughout this paper we use the Hubble constant $H_{\rm o}$= 70 km s$^{-1}$ Mpc$^{-1}$ and $\Omega_\Lambda=0.7$ and $\Omega_{\rm M} = 0.3$. At a distance of IC~5063, this results in 1 arcsec = 200 pc, (Wright 2006).}, selected because it represents one of the best examples  of  fast ($\sim 700$ \kms) outflow of neutral hydrogen and ionized gas (Morganti et al.\ 1998, 2007, see Fig.\ 1 for an overview)  that is located at about  0.5~kpc from the nucleus and  is coincident with the bright radio lobe  (Oosterloo et al.\ 2000). Therefore,  the mechanical
energy associated with the radio jet/lobe has been suggested to be the main mechanism that drives  the outflow. This object is, therefore, an ideal candidate for understanding the details of the gas associated with such interactions and its relevance  for feedback.

IC~5063 was previously observed and detected with SEST in CO(1-0) by Wiklind, Combes \& Henkel (1995). The availability of these data prompted us to apply for observations of the  CO(2-1) component. The  CO(1-0) profile reported by Wiklind et al. (1995) is, however, relatively narrow  ($\Delta v_{FWHM} = 163$ \kms) and offset in velocity compared, e.g. to the \HI\ emission profile.  In early-type galaxies, molecular and atomic neutral gas tends to show similar kinematics  (Davis et al. 2011). Thus, the new observations can also serve as a comparison for this puzzling difference.

\section{CO(2-1) in IC~5063: the APEX observations}

Observations with the Atacama Pathfinder EXperiment (APEX) 12-m antenna were conducted in two campaigns in
2008 using the APEX-1 instrument with the FFTS1 backend tuned to 230.538 GHz,  the frequency of the \co\ line at the redshift of IC~5063 ($z = 0.0110$).
The central position of IC~5063 (RA=20h52m02.3s Dec=-57d04m08s) was observed on August 8-9, 2008 in good weather conditions (for these frequencies). The precipitable water vapor (PWV) was $\leq 2$ mm. IC~5063 was observed for a total integration time of 146 minutes. 

The observations were made using 4096 channels, resulting in a velocity resolution of 0.32 km/s. The two fast Fourier transform
Spectrometer (FFTS) units almost fully overlap, resulting in a total bandwidth of 1~GHz ($\sim 1250$ \kms). For the individual scans of each of the FFTS units, a linear baseline was subtracted before adding all spectra. The final spectrum is smoothed to $\sim 15$ \kms\ bins. The rms-noise at 15 \kms\ resolution is $\sim 1$ mK. The data were reduced with the CLASS software from the Gildas package (http://www.iram.fr/IRAMFR/GILDAS). 

At the frequency of these observations, the spatial  resolution of APEX  is  $\sim 30^{\prime\prime}$ (i.e., covering a radius to $\sim 3$ kpc from the center). At this resolution, the central beam includes the whole radio continuum source  and the inner part of the gas disk observed in \HI\ emission (see Fig. 1 for reference).  Inside this radius,  the \HI\ emission covers the full velocity range of the rotating \HI\ disk (because of the steep rotation curve typical of early-type galaxies) and also includes the broad, blueshifted  component observed in \HI\ absorption against the radio source  (see Fig.\ref{fig:panel}).

\begin{figure}
\centerline{\psfig{file=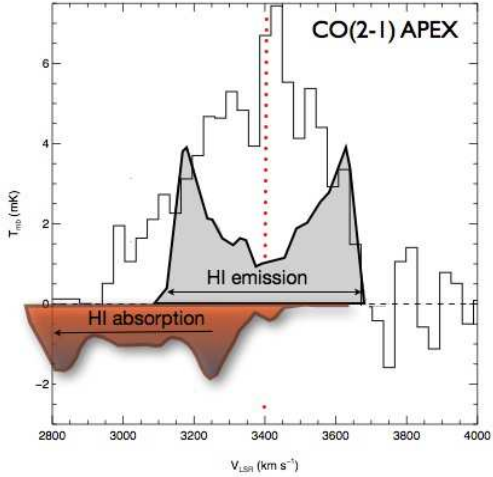,width=7cm,angle=0}}
\caption{Profile obtained from the APEX CO (2-1) observations. The profile is extremely broad (almost 800 km s$^{-1}$ FWZI) and asymmetric, showing a pronounced blueshifted ÒwingÓ. The profile reaches velocities well outside the range covered by  the regularly rotating large-scale gas disk. This is illustrated by the comparison with the velocities covered by the \HI\ emission (indicated in gray in the plot, with arbitrary intensity scale). The blueward wing of the \co) profile covers velocities similar to those of the \HI\ outflow detected in absorption.}
\label{fig:profile}
\end{figure}

The final \co\ profile observed in IC~5063  is shown in Fig. \ref{fig:profile}. 
The new result that is immediately apparent from this figure is the  large  width of the  \co\  profile:  about $800$ km/s FWZI.   Thus, the CO(2-1) profile appears to be significantly broader than what was derived in CO(1-0) by Wiklind et al. (1995).
The width and the central velocity  of the CO(2-1) line (3400 vs 3349 km/s in CO(1-0)) appears to be more consistent with the \HI\ profile (as can be seen in Fig. 2). Furthermore, the CO(1-0) does not appear to be asymmetric and does not have  a blueshifted wing. 

It is perhaps more interesting  that  the profile is also much broader than the integrated \HI\ emission profile obtained from the entire \HI\ disk. This is illustrated in Fig. 2 where the \HI\ emission  profile is taken from Morganti et al.\ (1998) and represents the full  velocity amplitude of the \HI\ gas in the large-scale disk, which is unaffected by the active galactic nucleus (AGN). Also indicated in Fig.\ 2 is the velocity range corresponding to the \HI\ outflow seen in absorption. The important result is  that the \co\ spectrum is asymmetric with a clear  {\sl blueshifted wing} that extends well outside the velocity range of the gas disk and into the velocity range of the \HI\ outflow.   The fact that the \co\ profile is broader than the \HI\ emission profile illustrates  that an extra component must be present, with velocities exceeding those of the gravitationally rotating gas. This suggests {\sl that the molecular gas is participating in the same outflow that drives out the \HI}. 

The bandwidth of the \co\ observations to some extent limits the sampling of the full range in velocities covered by the \HI\ absorption.  Therefore, the full width of the blueshifted component of the molecular gas is somewhat uncertain. However, the data clearly show that excess emission that  cannot be explained by poor subtraction of the linear baseline during the reduction steps.

\section{A new component of the gaseous outflow: the molecular gas} 

The blueshifted  wing in the \co\  profile suggests that in IC~5063 we are observing the molecular gas associated with an outflow in addition to the molecular gas associated with the large gas disk.  Despite the limitations present in the \co\ data mentioned above, we now derive some parameters to characterize the outflow of molecular gas. 

We estimated the flux of the outflow component by fitting two Gaussians to the CO profile. The result of the fitting is shown in Fig. 3.  In this way we obtained  $I_{\rm CO} = 0.54\pm 0.17$ K \kms. We used a  conversion factor of  39 Jy K$^{-1}$ for  {\sc APEX}. To estimate the H$_2$ masses we assumed  a luminosity ratio L$^{\prime}$CO(2-1)/L$^{\prime}$CO(1-0) = 1 typical of thermalized and optically thick gas, as could be the situation in  denser and hotter gas associated with an outflow. If this is the case, it could explain why the broad component was not detected  in Wiklind et al. (1995). However, the physical conditions of  these outflows are still poorly known and other possibilities have been proposed (Alatalo et al. 2011). 


The choice of the conversion factor $\alpha_X = M_{H_2}/L^{\prime}_{CO}$ M$_{\odot}$ (K \kms pc$^2$)$^{-1}$ is not easy because  very little is known about outflows (see discussion in Alatalo et al. 2011).
To be conservative, we used the two extreme values of  $\alpha_X = 0.8$ M$_{\odot}$ (K \kms pc$^2$)$^{-1}$  derived by   Downes \& Solomon (1998) for ultra-luminous infrared galaxies and the  standard $\alpha_X = 4.6$ M$_{\odot}$ (K \kms pc$^2$)$^{-1}$ (see Solomon \& Barrett 1991 for a discussion).  The derived mass of the molecular outflow ranges from $2.25\pm 0.70 \times 10^7$ \msun\  to $1.29\pm 0.40  \times 10^8$ \msun.
Thus, the derived mass of the molecular outflow appears to be  higher than that deduced for the \HI\ outflow (i.e., $3.6 \times 10^6$ \msun; Morganti et al. 2007). 

The  total mass  of the molecular gas - in the central beam -  following the regular rotation (i.e., centered on the systemic velocity) and  estimated using similar assumptions as described above, is  $4.96 \pm 0.50 \times 10^8$ \msun. 
This means that the ratio between the molecular masses derived from the CO(2-1) and the one quoted by Wiklind et al. (1995) and derived from the CO(1-0) is 1.3.

Following Alatalo et al. (2011), we  derived the mass outflow rate of the molecular gas. The spatial resolution of the {\sc APEX}  observations do not allow one to identify the exact location of the molecular outflow in IC~5063. 
If the molecular outflow is located in/near the western and radio-bright lobe, i.e., about 0.5 kpc away from the nucleus, the time for the outflow to reach this radius  is $\tau_{dyn} \sim 1$ Myr, assuming the highest velocity observed for the CO ($\sim 400$ \kms\ relative to the systemic velocity). 
The mass outflow rate estimated as $\dot M = M/\tau_{dyn}$ has a lower limit of $\sim 22$  \msunyr (for the  conservative assumption of  $\alpha_X = 0.8$ \msun), reaching up to 129 \msunyr\ using the result of the Gaussian fitting and $\alpha_X = 4.6$ \msun.  

For these relatively high mass outflow rates, one may wonder whether they can be easily supported in the case of IC~5063.  The  total mass  of the molecular gas - in the central beam - that  follows the regular rotation is $4.96 \pm 0.50 \times 10^8$ \msun\  (based on the results of the Gaussian fitting and using $\alpha_X = 4.6$ \msun). 
Of this, we can assume that about a tenth ($\sim 5 \times 10^7$ \msun) is included in the central kpc. 
This is derived by  very roughly assuming that it follows the ratio between the beam area (radius 3 kpc) and the area affected by the jet (radius $\sim 1$ kpc). Due to the collimated nature of the outflow,  only part of the gas in the inner region is affected by the jet at any instant. However, because of the rotation of the inner gas disk, eventually  the entire gas of the inner disk will be affected. The orbital timescale in the inner region is about $3 \times 10^7$  years, implying an average mass outflow rate of $\sim  1$ \msunyr. This is, of course, a very simplified view, and the mass of the molecular gas in the central beam  could be higher if an exponential disk distribution were assumed. However, even for a  substantially larger portion of molecular gas, the mass outflow rate would stay below what was measured. This difference may suggest  that the gas outflow in IC~5063 occurs  in bursts and, at present,  is in a particularly strong phase. If this is the case, this would have interesting implications for an explanation of some of the peculiar characteristics of this source, and we  return to this question in Sec. 5.

\section{Origin of the outflow}

Existing data on the observed outflow of ionized and atomic gas in IC~5063 suggest that these outflows are driven by the interaction between the radio jet and the ISM.  The  molecular  component could be part of the same outflow. 

Molecular gas associated with the jet/ISM  interaction has also been suggested on the basis of observations of  warm molecular gas by {\sc NICMOS}  (Kulkarni et al. 1998) and, more recently, in the mid-IR by  {\it Spitzer}  (Guillard et al.\ 2012).
Kulkarni et al.\ (1998) found that the [FeII] $\lambda {\rm 1.644}~\mu$m and H$_2$ $\lambda {\rm 2.1218}~\mu$m emission presents a one-to-one spatial correspondence with the  radio lobes, suggesting that these lines originate in shocks produced by the advancing radio jets. Furthermore, the asymmetry in the [FeII]/H$_2$ ratio between the eastern and western  sides suggests an excess of molecular gas on the W side, where the radio lobe is brighter. This reinforces the idea of the jet interacting with a particularly rich and dense medium.  
More recently, this scenario has obtained support from {\it Spitzer} mid-IR data. The warm H$_2$ appears to be particularly bright in radio galaxies with an \HI\ outflow, and most likely  the dissipation of the  kinetic energy of the radio jet is powering this bright H$_2$ emission. IC~5063 is one of such cases (Guillard et al.\ 2012). 
As described in Guillard et al.\ (2012), and following theoretical studies (Mellema et al.\ 2002, Gaibler et al.\ 2012, Wagner \& Bicknell  2011, Wagner et al. 2012),  the jet itself cannot affect a significant fraction of the ISM of the host galaxy because it is very collimated. Instead, the expanding cocoon of hot and tenuous gas inflated by the jet (e.g.\ Begelman \& Cioffi 1989) could transfer part of its kinetic energy to the surrounding ISM by driving shocks and/or by turbulent mixing.
Numerical simulations have studied the effects of a cocoon of shocked material running over a clumpy medium and/or a massive molecular cloud  (Mellema et al.\ 2002, Gaibler et al.\ 2012, Wagner \& Bicknell  2011, Wagner et al. 2012). A compression phase of this pre-existing gas is expected, followed by fragmentation and cooling. This process generates dense, cool, fragmented structures that can be entrained and accelerated by the cocoon that surrounds the radio jets/lobes. The key to  these models is that radiative cooling must be very efficient  to be able to produce accelerated clouds of cold gas (which would be observable in \HI\  and CO). In this scenario, \HI\ would represent an intermediate phase in the cooling process of the warm gas, and the molecular gas  would be the final stage.  The simulations also showed that the transfer efficiency of kinetic energy and momentum from the jet to the dense ($n_H > 10^2$ cm$^{-3}$) gas can be high (10-70\%).

\begin{figure}
\centerline{\psfig{file=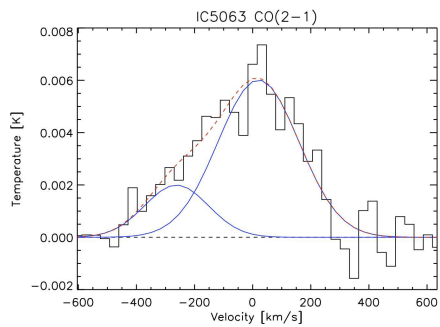,width=6cm,angle=0}}
\caption{Fit of the \co\ profile showing two Gaussian components. See text for details.}
\label{fig:GausFit}
\end{figure}

In summary, the detection of cold gas and the expectations of the simulation underline the complex and multi-phase nature of the outflows. 

\section{Discussion and conclusions}

We have presented the detection of  molecular gas outflow in the radio-loud Seyfert galaxy IC~5063. This represents another case - in addition to Mrk~231 (Feruglio et al.\ 2010), NGC~1266  (Alatalo et al.\ 2011),   and 4C12.50 (Dasyra \& Combes 2012) - where fast outflows of  molecular, atomic neutral  and   ionized gas are found to co-exist in the same object. 

In addition to the possibility that these  outflows are driven by the interaction between the radio jet plasma and the rich ISM, it is also interesting to consider the possibility that the same interaction has a major impact also on the characteristics of the radio source and, in particular, on enhancing the radio emission. This could also explain some of the intriguing characteristics of IC~5063  described above (i.e., high radio power for a Seyfert galaxy, large asymmetry between the two radio lobes, high mass outflow rate at the present stage).
Because of the compression of the magnetic field and the increased density of particles, the radio luminosity can be boosted by the interaction (Tadhunter et al. 2011). This has been suggested for the young, compact source PKS~1814-637 hosted by a disk galaxy with a rich ISM (i.e., H$_2$, and PAH emission  along with \HI\ and silicate absorption features) --- characteristics that are quite rare among powerful radio sources.  Because of this,  PKS~1814-637 and other similar objects have been suggested by Morganti et al. (2011) to be something like {\sl ÓimpostorsÓ}: objects intrinsically of low power  that are selected in  radio-flux-limited samples because of the  efficient conversion of jet power into radio emission. These are objects with a rich, dense ISM that are expected to have unusually strong interactions between the jets and the ISM (Tadhunter et al. 2011). 
They may also represent a missing link between radio galaxies and radio-loud Seyfert galaxies. IC~5063 is one of the most radio-loud Seyfert galaxies. Thus, it is intriguing that in this object we see clear evidence that  the radio jet strongly interacts with the ISM, in agreement with what is expected in the above scenario. The suggestion that the mass outflow rate is, at present, particularly high in IC~5063,  additionally supports  that this object is in a special phase at the moment.

Finally, although it is massive, the cold gas component of the AGN-driven outflows does not always  appear  to be large enough to match what is required by quasar-feedback models (di Matteo et al. 2005).   In  IC~5063, the kinetic power derived by combining the \HI\ and CO components (assuming they are tracing the same outflow)  is between  $7.8 \times 10^{42}$ and $1.7 \times 10^{43}$ \ergs, depending on the mass outflow rate assumed for the molecular gas. 
These estimates correspond to  $\sim 10 - 20$ \% of the nuclear AGN bolometric luminosity (as derived in Morganti et al. 2007), which   suggests that the impact of the outflow on the ISM is relevant in the evolution of the host galaxy. However, one should consider the possibility that  the mass outflow rate is particularly high at present but cannot be maintained at this level for  a long time, and therefore may have an effect only for a limited period in the life of the galaxy.  

The next step in the study of this object is to use higher resolution and more sensitive observations to find the exact location with respect to the other components of the molecular outflow, and investigate its characteristics in greater detail.  In this way, it will be possible  to verify  the scenario sketched here.

\begin{acknowledgements}
We would like to thank Francoise Combes and  the referee for a number of valuable suggestions that  greatly  improved the quality of this paper. This publication is based on data acquired with the Atacama Pathfinder Experiment (APEX). APEX is a collaboration between the Max-Planck-Institut fur Radioastronomie, the ESO, and the Onsala Space Observatory. We acknowledge the use of GILDAS software for the  reduction of the APEX  data. 

\end{acknowledgements}

\end{document}